# Acoustic Lenses Design based on the Rays Inserting Method

**Liuxian Zhao[1,2], Chuanxing Bi[1], Miao Yu[2,3,*]**

[1]Institute of Sound and Vibration Research, Hefei University of Technology, 193 Tunxi Road, Hefei 230009, China

[2]Institute for Systems Research, University of Maryland, College Park, MD, 20742, USA

[3]Department of Mechanical Engineering, University of Maryland, College Park, Maryland 20742, USA

*Author to whom correspondence should be addressed: mmyu@umd.edu

**ABSTRACT**

The ability to control and manipulate elastic waves is important for applications such as structural health monitoring, signal processing, and vibration isolations. In this paper, we investigated the feasibility of using the Rays Inserting Method (RIM), an approach originally proposed for optical elements, to design structural components for flexural wave manipulation. The RIM entails a simple process that allows to design thickness variations in a thin plate with a desirable refractive index distribution for an intended wave path. Based on this method, a focusing/collimating lens and a waveguide that rotates the wavefront by 45° were designed and studied. Frequency domain simulations and time-based experimental characterizations were carried out. The results demonstrated that the effectiveness of the RIM for designing variable thickness structures for manipulation of flexural wave propagation along desired paths.

**1. Introduction**

The ability to control and manipulate the propagation of elastic waves is important for many applications such as structural health monitoring [1], vibration isolation [2], telecommunication [3], signal processing [4, 5], and wave cloaking [6-8], etc. The wave propagation paths can be modulated to produce rotation [9], focusing [10-12], collimation [13, 14], cloaking [7, 8] and splitting [15], amongst other effects[16]. A common method of elastic wave manipulation is to employ periodic structures to alter the effective refractive index of the medium through Bragg scattering or local resonances [17]. Refractive index modifications can also be made through variations in physical properties of the medium, for example, the thickness of a thin plate [18, 19]. The latter method offers the advantage of broadband performance [9], which can be important for applications such as those involving Lamb waves.

To design a medium with desired elastic wave control properties (e.g., collimation, splitting, bending), the spatial distribution of refractive index must be obtained, which is not always straightforward. The transformation methods [20, 21] that make use of the form-invariance of Maxwell and acoustic equations to design optical and acoustic lenses and waveguides cannot be directly applied for elastic wave control due to the form-dependence of elastodynamic equations, unless the wave propagation in the medium possesses a longitudinal wave velocity that is much higher than its transverse wave velocity [22]. To solve this problem, conformal transformation acoustics or quasi-conformal transformation acoustics were usually used [23, 24]. For example, Chen *et al.* [25] designed an elastic metamaterial based on the conformal transformation acoustics method for guiding the flexural wave propagation in a thin plate structure. However, the complex and anisotropic material properties still pose challenges for these methods [26, 27]. In order to overcome the challenges encountered by using the conformal transformation acoustics or quasi-conformal transformation acoustics methods, we propose to use Rays Inserting Method (RIM) for designing elastic wave lenses.

The RIM is a technique that was originally proposed by Taskhiri and Amirhosseini to design dielectric optical devices [28]. With the RIM, rays are inserted between two ends of the optical device to be designed and the refractive indices of the points along the rays are obtained. The RIM was initially used to design three dielectric optical devices: power splitter, bend, and flat lens [28]. A In a subsequent study, a broadband hemispherical optical wave collimator lens was designed by using the RIM, which can convert a point source to a plane wave [29]. This work is the first attempt to use the RIM to design elastic wave lenses.

## 2. Rays Inserting Method (RIM)

In this work, we investigate the feasibility of using the RIM for the design of elastic wave lenses. In the two-dimensional (2D) cartesian coordinates, an inserted ray starts from a point ($x_0$, $y_0$) with a refractive index of $n_0$ and an angle of $α_0$, and ends at a point ($x_s$, $y_s$) with a refractive index of $n_s$ and an angle of -$α_s$, as shown in Figure 1. The trajectory of a desired wave path, $\vec{r}(t)$, is first inserted between the incident and exit edges of a 2D lens according to:

$$\vec{r}(t) = x(t)\widehat{a_x} + y(t)\widehat{a_y}, \qquad (1)$$

where $\widehat{a_x}$ and $\widehat{a_y}$ are the unit vectors in the $x$ and $y$ directions, $t$ is an independent parametric variable that varies from 0 at the incident edge to $t_s$ at the exit edge. $x(t)$ and $y(t)$ are provided in the supplementary material.

The refractive index along the inserted ray can be obtained by using the Eikonal equation [30] with detailed information in the supplementary.

$$n^2(x, y, z) = \left(\frac{dx}{dt}\right)^2 + \left(\frac{dy}{dt}\right)^2, \qquad (2)$$

that is:

$$n^2(t) = \left[n_0 \cos(\alpha_0) \cos\left(p\frac{t}{t_s}\right) + \frac{A}{t_s}\sin\left(p\frac{t}{t_s}\right)\right]^2 + \left[n_0 \sin(\alpha_0) \cos\left(q\frac{t}{t_s}\right) + \frac{B}{t_s}\sin\left(q\frac{t}{t_s}\right)\right]^2.$$

(3)

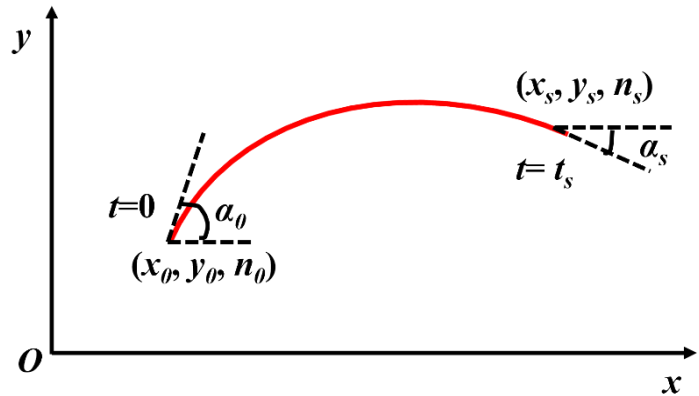

**Figure 1: Schematic diagram showing a 2D wave path starting from an incident point, with coordinates of ($x_0$, $y_0$), refractive index of $n_0$, and angle of approach $a_0$, to the exit point with coordinates of ($x_s$, $y_s$), refractive index of $n_s$, and angle of approach -$a_s$. $t$ is an independent variable that varies from 0 at the incident point to $t_s$ at the exit point.**

For flexural wave propagation through a thin plate, the phase velocity, $c_p$, is a function of the plate thickness, $h$, which can be expressed as $c_p = (\frac{Eh^2\omega^2}{12\rho(1-\nu^2)})^{\frac{1}{4}}$. Here, $\omega$ is the angular frequency, $E$ is the Young's modulus, $\rho$ is the density, and $\nu$ is the Poisson ratio [10]. According to Snell's law, the relationship between the refractive index, $n$, and the phase velocity is $= \frac{c_0}{c_p}$, where $c_0$ refers to the phase velocity of the flexural wave propagation through a thin plate with a constant thickness of $h_0$. Therefore, the refractive index of each spatial point in an elastic wave lens or waveguide obtained by using Eq. (3) can be realized by designing the plate thickness of that point according to:

$$n = \sqrt{\frac{h_0}{h}} \qquad (4)$$

To illustrate the procedures of utilizing the RIM approach, we first plot the wave path trajectories of a flexural wave lens (Figure 2a), as well as a waveguide (Figure 2b). The lens was designed to act as a collimator for a point source, or a focusing lens for plane waves for waves traveling in the opposite direction. In the case of collimation, the wave paths start from a point source at $(x_0, y_0) = (0, 0)$ with different incident angles of $\alpha_0$, and the exit paths are located at $x_s = 0.04$ m with different $y_s$ values and an angle of $\alpha_s = 0°$ (Figure 2a). The waveguide, on the other hand, was designed to rotate the direction of wave propagation by 45° (Figure 2b). The paths start from the incident edge at $x_0 = 0$ m and $y_0$ varying from 0.04 m to 0.08 m with an angle of $\alpha_0 = 0°$, and the paths exit with an angle of $\alpha_s = 45°$.

Using these desired trajectories, the refractive index distributions were calculated based on Eq. (3) and plotted in Figures 2c and 2d. Using Eq. (4), the thickness profiles of the lens (Figure 2e) and waveguide (Figure 2f) defined in a thin plate can then be obtained. Note that the refractive indices $n_0$ and $n_s$ at the incident and exit edges of the lens and waveguide are required to be the same as those of the background media in order to achieve impedance matching and reduce reflections. This implies that the thickness of plate at the incident and exit edges of the lens or waveguide should be the same as the constant plate thickness $h_0$ of the overall plate.

Note that only the refractive indices along the inserted rays can be calculated by using the RIM. The refractive indices at other locations were obtained by using interpolation with the MATLAB software. The more rays inserted between the input and output edges, the more accurate and smoother the distribution of the obtained refractive index will be.

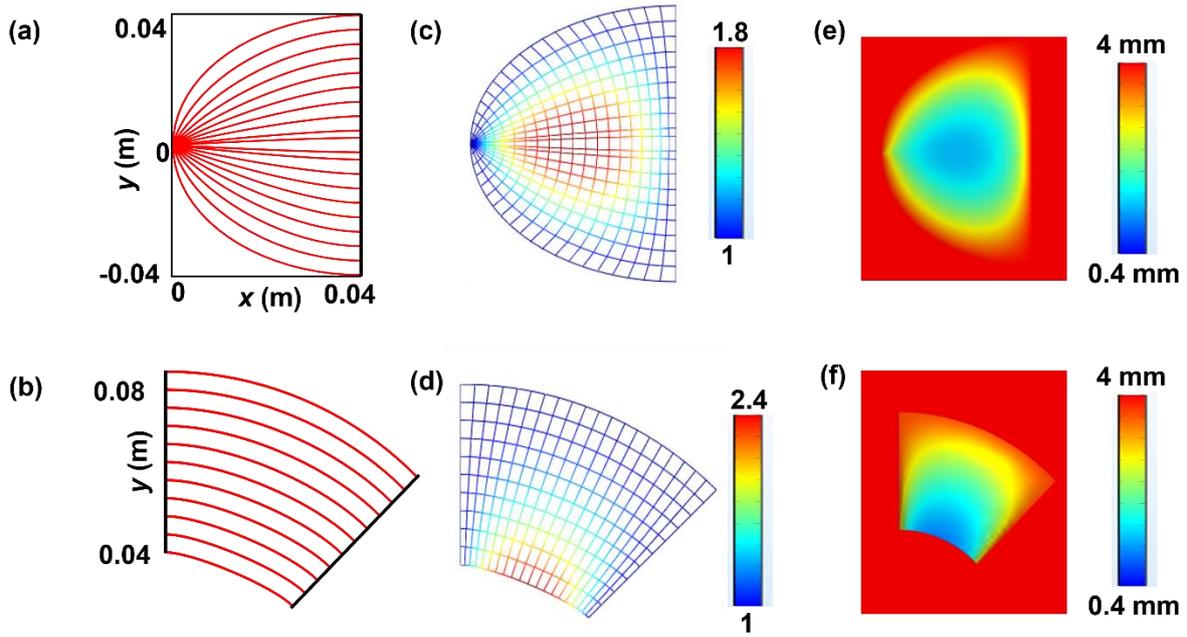

**Figure 2:** Desired flexural wave path trajectories of (a) lens and (b) waveguide. Corresponding refractive index distributions of (c) lens and (d) waveguide. Thickness variation profiles of (e) lens and (f) waveguide in a thin plate.

## 3. Numerical Simulations

To validate the lens and waveguide designs, numerical simulations in COMSOL were carried out for the structures with thickness profiles obtained in Figures 2e and 2f. In the simulations, the lens and waveguide structures were defined in a thin aluminium plate (thickness of 4 mm, mass density $\rho = 2700$ kg/m$^3$, Young's modulus $E = 70$ GPa, and Poisson's ratio $v = 0.33$). Perfectly Matched Layer (PML) boundary condition was used to reduce the boundary reflections for frequency domain analyses. For the flexural wave collimator, a point source located at (0, 0) with a frequency of 60 kHz was used for excitation. For the flexural waveguide, a line source located at $x = 0$ m and $y$ ranging from 0.04 m to 0.08 m with a frequency of 60 kHz was used. The full field wave propagations for the collimator lens and waveguide are plotted in Figures 3a and 3b, respectively. The results clearly demonstrated the intended wave manipulation paths.

Furthermore, more complex wave trajectories realized by using the combination of multiple elements were investigated. The following scenarios for manipulation of an input plane wave were studied in the simulations: i) focus and collimation with two cascaded lenses (Figure 3c), ii) wave guiding in a S-shaped trajectory (Figure 3d), iii) wave guiding in a U-shaped trajectory (Figure 3e), and iv) waveguiding in a double-S-shaped trajectory (Figure 3f). Defects introduced in the simulations (white blocks in Figures 3 (c)-(f)) were found to have negligible effect on the wave guiding performance of the structures. These results validated the effectiveness of achieving various wave manipulation structures through the variable thickness cavities in a thin plate designed by using the RIM. Furthermore, these results demonstrated that the RIM can be used to design wave guiding structures with obstacle avoidance, cloaking, and isolation properties.

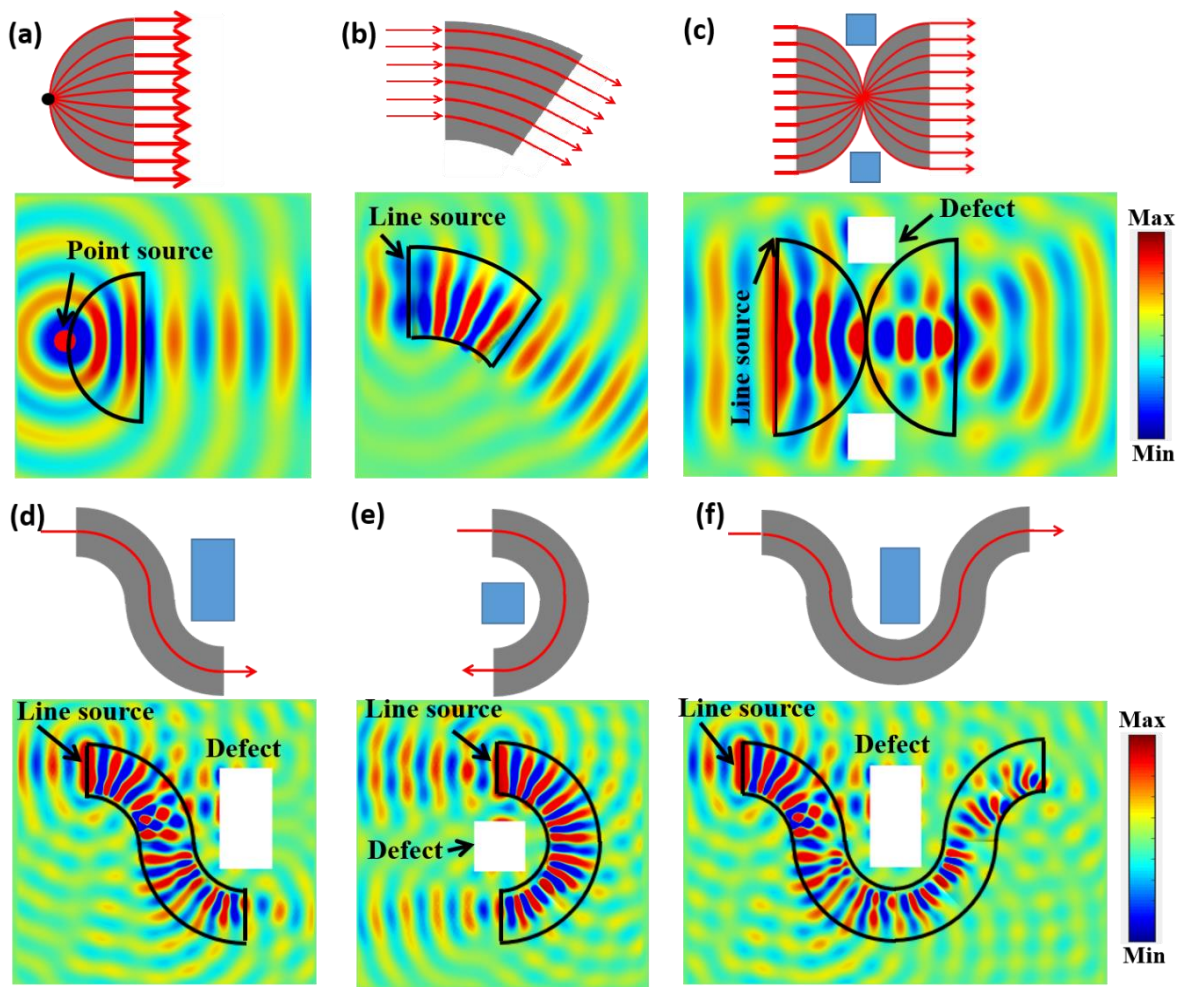

**Figure 3:** Schematic diagrams (top) and corresponding simulation results (bottom) for (a) collimation lens, (b) waveguide that rotates wavefront by 45°, (c) waveguide combining a focusing and a collimation lens, (d) S-shaped waveguide composed of four 45° waveguides, (e) U-shaped waveguide composed of four 45° waveguides to rotate wavefront by 180°, (f) double S-shaped waveguide composed of eight 45° waveguides. The black line in the simulation results represents the outline of the lens or waveguide. The colour bar indicates out-of-plane displacement. Defects are represented by blue boxes in the schematic diagrams and white boxes in the simulation results.

## 4. Experimental Measurements

Experimental studies were also carried out to validate the performance of the collimation lens and the 45° waveguide. The experimental setup and the fabricated variable thickness structures are shown in Figure 4. The lens and waveguide were each fabricated on a thin aluminium plate (aluminium 6061, McMaster-Carr) with dimensions of 0.3 m × 0.3 m× 0.004 m. All four edges of the plates were covered with clay to reduce reflections from boundaries. A circular piezoelectric disc (STEMiNC Corp.) (12 mm in diameter and 0.6 mm in thickness) was used as a point source to generate a circular waveform for the collimator lens (Figure 4a). Two rectangular piezoelectric transducers (STEMiNC Corp.) with dimensions of 20 mm× 15 mm × 1 mm were used to generate a plane wave for the waveguide (Figure 4b). The piezoelectric transducers were bonded to the plate by using an adhesive (2P-10 from Fastcap, LLC). During the experiments, the plates were fixed on a vertical frame. A scanning laser Doppler vibrometer (Polytec PSV-400) was used to collect the propagating wave field by recording the out-of-plane component of the particle velocity on the flat surface of the plate, as shown in Figure 4(c).

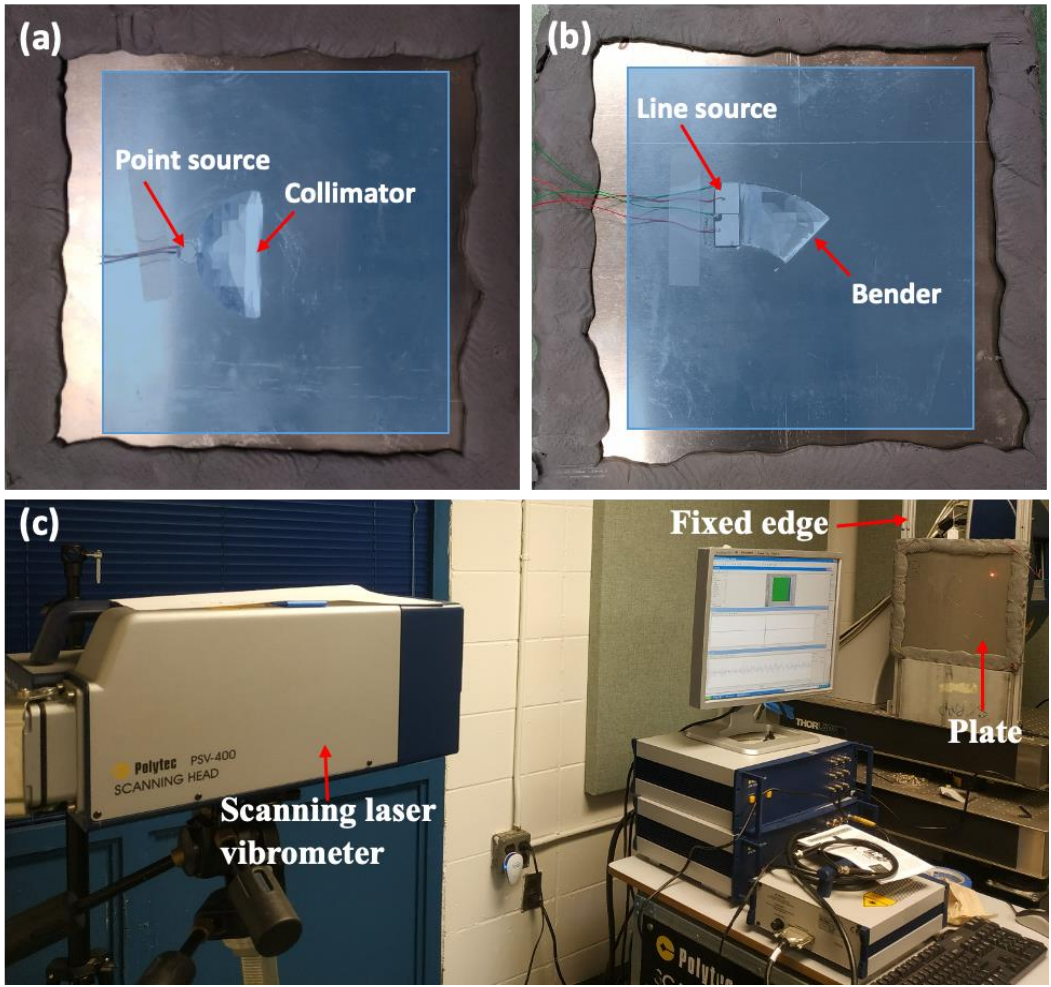

**Figure 4: Fabricated (a) collimation lens and (b) waveguide. (c) A photo of the experimental setup. A scanning laser vibrometer was used to measure the full-field wave propagation on the flat side of the plate in the vicinity of the cavities. The boundaries of the thin plate were covered with clay and constrained by two vertical fixtures.**

In Figure 5, the experimental results at different time steps are shown. For the collimation lens, a point source with 3-count tone bursts at a center frequency of $f = 60$ kHz was used for excitation at the same location as that used in the numerical simulations. The input amplitude of the signal was 1 V and it was pre-amplified 30 times using an amplifier. The displacement fields of wave propagation at different time instances of 0.02 ms, 0.05 ms and 0.08 ms are shown in Figures 5a - 5c. From 0 ms to 0.02 ms, the generated circular waveform travelled through the wave collimator and the wavefront became flatter in the process. At the time instance of 0.05 ms, the propagating wave exited the lens as a plane wave. By 0.08 ms, the flexural wave left the lens completely and propagated forward.

For the 45° waveguide, a line source along *y* direction was used for excitation at the same location as that used in the numerical simulations with a central frequency of 60 kHz. Similar to the analysis for the collimation lens, the full field wave propagations at different time instances, 0.02 ms, 0.05 ms, and 0.08 ms are shown in Figures 5d - 5f. From 0 ms to 0.02 ms, the plane flexural wave propagated from the line source and began interacting with the waveguide. At the time instance of 0.05 ms, the flexural wave travelled through the lens and began exiting. At the time instance of 0.08 ms, the plane wave was observed to propagate forward at an angle of 45º with respect to the original travel direction. These experimental results agree well with the simulation results and demonstrate that the variable thickness profiles designed for the collimator and 45° waveguide can be used to successfully manipulate the flexural waves as the intended designs.

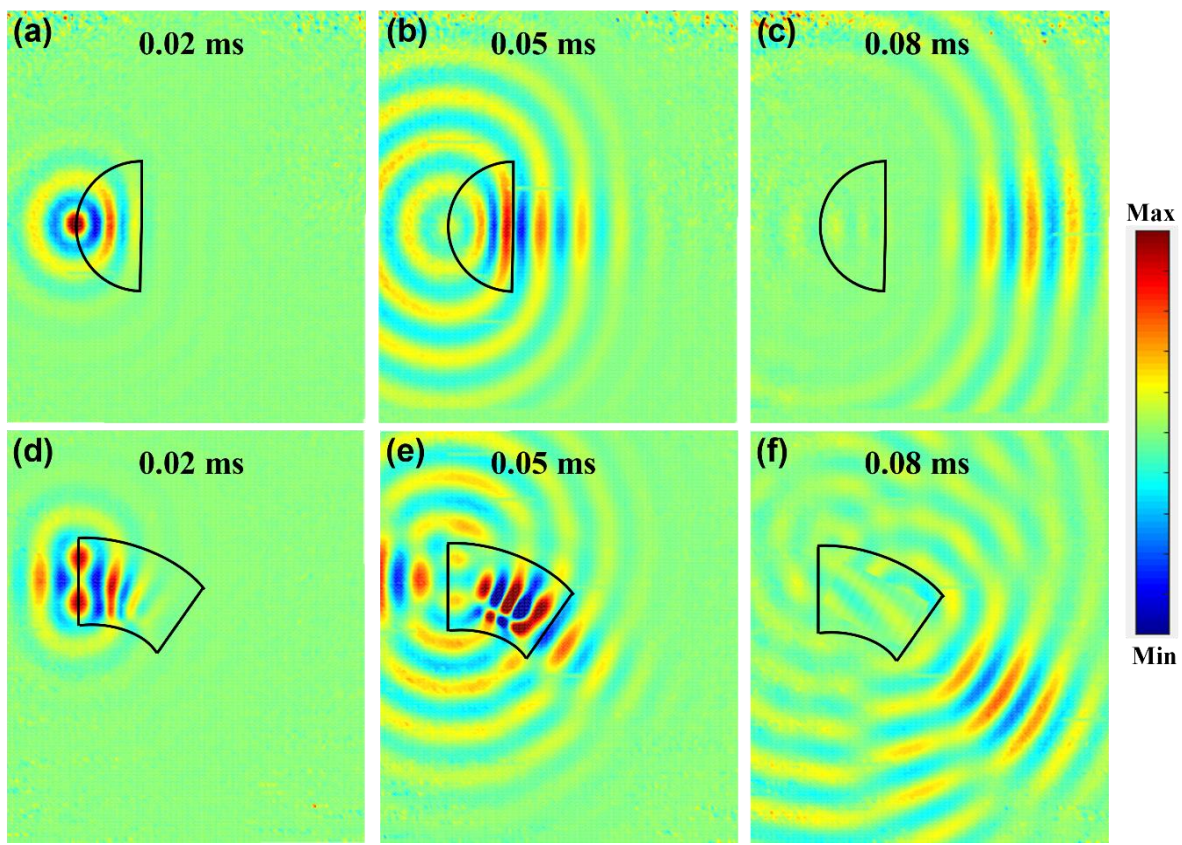

**Figure 5: Experimentally measured transient responses for collimation lens and waveguide. Full field wave propagation results for the collimator lens at time instances of (a) 0.02 ms, (b) 0.05 ms, and (c) 0.08 ms. Full field wave propagation results for the 45° waveguide at time instances of (d) 0.02 ms, (e) 0.05 ms, and (f) 0.08 ms. The black outlines indicate the lens and waveguide. The colour bar indicates the out-of-plane displacement.**

## 5. Conclusions

In conclusion, we have demonstrated, through simulations and experiments, that the Rays Inserting Method (RIM) is a viable technique for designing various structures (e.g., lenses and waveguides) for controlling flexural wave propagation in a thin plate. The RIM starts with the introduction of desired wave path trajectories, from which the refractive index distribution of a structural element can be calculated. The refractive index distribution, in turn, can be realized by using a variable thickness structure and its thickness profile can be obtained through Snell's law. Combining several variable thickness structure elements, complex wave trajectories can be realized for cloaking, isolation, and obstacle avoidance purposes.

**Conflict of Interest**

The authors declare no conflict of interest.

**Data Access Statement**

The data that support the findings of this study are available from the corresponding author upon reasonable request.

# References

1. Rizzo, P., J.-G. Han, and X.-L. Ni, *Structural Health Monitoring of Immersed Structures by Means of Guided Ultrasonic Waves.* Journal of Intelligent Material Systems and Structures, 2010. **21**(14): p. 1397-1407.
2. Zhao, L., *Low-frequency vibration reduction using a sandwich plate with periodically embedded acoustic black holes.* Journal of Sound and Vibration, 2019. **441**: p. 165-171.
3. Vasseur, J.O., et al., *Waveguiding in two-dimensional piezoelectric phononic crystal plates.* Journal of Applied Physics, 2007. **101**(11): p. 114904.
4. Brun, M., I.S. Jones, and A.B. Movchan, *Vortex-type elastic structured media and dynamic shielding.* Proceedings of the Royal Society A: Mathematical, Physical and Engineering Sciences, 2012. **468**(2146): p. 3027-3046.
5. Cha, J., K.W. Kim, and C. Daraio, *Experimental realization of on-chip topological nanoelectromechanical metamaterials.* Nature, 2018. **564**(7735): p. 229-233.
6. Ning, L., Y.-Z. Wang, and Y.-S. Wang, *Active control cloak of the elastic wave metamaterial.* International Journal of Solids and Structures, 2020. **202**: p. 126-135.
7. Zhao, L. and M. Yu, *Structural Luneburg lens for broadband cloaking and wave guiding.* Scientific Reports, 2020. **10**(1): p. 14556.
8. Sklan, S.R., R.Y.S. Pak, and B. Li, *Seismic invisibility: elastic wave cloaking via symmetrized transformation media.* New Journal of Physics, 2018. **20**(6): p. 063013.
9. Darabi, A., et al., *Broadband Bending of Flexural Waves: Acoustic Shapes and Patterns.* Scientific Reports, 2018. **8**(1): p. 11219.
10. Zhao, L., C. Lai, and M. Yu, *Modified structural Luneburg lens for broadband focusing and collimation.* Mechanical Systems and Signal Processing, 2020. **144**: p. 106868.
11. Zhao, L., H. Kim, and M. Yu, *Structural Luneburg lens for broadband ultralong subwavelength focusing.* Mechanical Systems and Signal Processing, 2023. **182**: p. 109561.
12. Zhao, L., T. Horiuchi, and M. Yu, *Broadband ultra-long acoustic jet based on double-foci Luneburg lens.* JASA Express Letters, 2021. **1**(11): p. 114001.
13. Ma, P.S., H.J. Lee, and Y.Y. Kim, *Dispersion suppression of guided elastic waves by anisotropic metamaterial.* The Journal of the Acoustical Society of America, 2015. **138**(1): p. EL77-EL82.
14. Zhao, L., T. Horiuchi, and M. Yu, *Broadband acoustic collimation and focusing using reduced aberration acoustic Luneburg lens.* Journal of Applied Physics, 2021. **130**(21): p. 214901.
15. Jin, Y., et al., *Multimodal and omnidirectional beam splitters for Lamb modes in elastic plates.* AIP Advances, 2016. **6**(12): p. 121602.
16. Zhao, L., T. Horiuchi, and M. Yu, *Acoustic waveguide based on cascaded Luneburg lens.* JASA Express Letters, 2022. **2**(2): p. 024002.
17. Ma, G. and P. Sheng, *Acoustic metamaterials: From local resonances to broad horizons.* Science Advances, 2016. **2**(2): p. e1501595.
18. Climente, A., D. Torrent, and J. Sánchez-Dehesa, *Gradient index lenses for flexural waves based on thickness variations.* Applied Physics Letters, 2014. **105**(6): p. 064101.
19. Lee, D., et al., *Singular Lenses for Flexural Waves on Elastic Thin Curved Plates.* Physical Review Applied, 2021. **15**(3): p. 034039.
20. Chen, H. and C.T. Chan, *Acoustic cloaking and transformation acoustics.* Journal of Physics D: Applied Physics, 2010. **43**(11): p. 113001.
21. Popa, B.-I., L. Zigoneanu, and S.A. Cummer, *Experimental Acoustic Ground Cloak in Air.* Physical Review Letters, 2011. **106**(25): p. 253901.
22. Gao, H. and Z. Xiang, *Manipulating Elastic Waves with Conventional Isotropic Materials.* Physical Review Applied, 2019. **11**(6): p. 064040.
23. Dong, E., et al., *Bioinspired Conformal Transformation Acoustics.* Physical Review Applied, 2020. **13**(2): p. 024002.

24. Zhao, L., et al., *Ultrasound beam steering with flattened acoustic metamaterial Luneburg lens.* Applied Physics Letters, 2020. **116**(7): p. 071902.
25. Chen, Y., J. Hu, and G. Huang, *A design of active elastic metamaterials for control of flexural waves using the transformation method.* Journal of Intelligent Material Systems and Structures, 2015. **27**(10): p. 1337-1347.
26. Liang, L. and S.V. Hum, *Wide-angle scannable reflector design using conformal transformation optics.* Optics Express, 2013. **21**(2): p. 2133-2146.
27. Mei, Z.L., J. Bai, and T.J. Cui, *Experimental verification of a broadband planar focusing antenna based on transformation optics.* New Journal of Physics, 2011. **13**(6): p. 063028.
28. Taskhiri, M.M. and M. Khalaj Amirhosseini, *Rays inserting method (RIM) to design dielectric optical devices.* Optics Communications, 2017. **383**: p. 561-565.
29. Taskhiri, M.M. and M.K. Amirhosseini, *Design of a broadband hemispherical wave collimator lens using the ray inserting method.* Journal of the Optical Society of America A, 2017. **34**(7): p. 1265-1271.
30. González-Acuña, R.G. and H.A. Chaparro-Romo, *The eikonal equation*, in *Stigmatic Optics*. 2020, IOP Publishing. p. 2-1-2-13.